\documentclass{article}

\usepackage{arxiv}

\usepackage[utf8]{inputenc} 
\usepackage[T1]{fontenc}    
\usepackage{hyperref}       
\usepackage{url}            
\usepackage{booktabs}       
\usepackage{amsfonts}       
\usepackage{nicefrac}       
\usepackage{microtype}      
\usepackage{lipsum}		
\usepackage{graphicx}
\usepackage{natbib}
\usepackage{doi}

\title{From Code Generation to Software Testing: \\ AI Copilot with Context-Based RAG}

\author{ \href{https://orcid.org/0000-0002-5183-1248}{\includegraphics[scale=0.06]{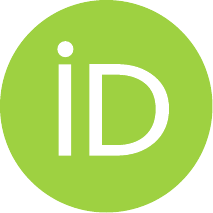}\hspace{1mm}Yuchen Wang} \\
	Nanyang Technological University\\
	Singapore \\
	\texttt{yuchen011@e.ntu.edu.sg} \\
	\And
	\href{https://orcid.org/0009-0004-0779-7179}{\includegraphics[scale=0.06]{orcid.pdf}\hspace{1mm}Shangxin Guo} \\
	City University of Hong Kong\\
	Hong Kong \\
	\texttt{sxguo2-c@my.cityu.edu.hk} \\
    \And
	\href{https://orcid.org/0000-0002-6624-9752}{\includegraphics[scale=0.06]{orcid.pdf}\hspace{1mm}Chee Wei Tan} \\
	Nanyang Technological University\\
	Singapore \\
	\texttt{cheewei.tan@ntu.edu.sg} \\
}

\date{}

\renewcommand{\shorttitle}{\textit{arXiv} Template}

\hypersetup{
pdftitle={A template for the arxiv style},
pdfsubject={q-bio.NC, q-bio.QM},
pdfauthor={David S.~Hippocampus, Elias D.~Striatum},
pdfkeywords={First keyword, Second keyword, More},
}

\begin{document}
\maketitle

\begin{abstract}
The rapid pace of large-scale software development places increasing demands on traditional testing methodologies, often leading to bottlenecks in efficiency, accuracy, and coverage. We propose a novel perspective on software testing by positing ``bug detection" and ``coding with fewer bugs" as two interconnected problems that share a common goal—reducing bugs with limited resources. We extend our previous work on AI-assisted programming, which supports code auto-completion and chatbot-powered Q\&A, to the realm of software testing. We introduce ``Copilot for Testing," an automated testing system that synchronizes bug detection with codebase updates, leveraging context-based Retrieval Augmented Generation (RAG) to enhance the capabilities of large language models (LLMs). Our evaluation demonstrates a 31.2\% improvement in bug detection accuracy, a 12.6\% increase in critical test coverage, and a 10.5\% higher user acceptance rate, highlighting the transformative potential of AI-driven technologies in modern software development practices.
\end{abstract}

\keywords{AI-assisted Testing \and Retrieval-Augmented Generation (RAG) \and Software Quality Assurance \and Bug Detection \and Automated Software Testing}

\section{Introduction}
In the evolving landscape of software development, the rapid pace of innovation and the increasing complexity of systems demand equally sophisticated tools to ensure reliability and efficiency. Traditional testing methods, while often reliant on manual efforts supplemented by semi-automated tools, often fall short in addressing the dual challenges of identifying bugs effectively and minimizing the bug rate in generated code. This gap not only impedes the development process but also affects the overall software quality, leading to potential oversights and inadequate test coverage \cite{k1, k2, k3}.

The industry's shift towards continuous integration and deployment accentuates the need for more efficient testing processes. In this context, artificial intelligence (AI) emerges as a game-changer for enhancing software testing. Recent advancements in computational models, particularly large language models (LLMs) empowered by retrieval-augmented generation (RAG), offer new avenues for improving testing methods. These AI-driven technologies can analyze extensive codebases and detect complex patterns that may escape traditional testing approaches, thereby enabling the generation of precise and comprehensive test cases\cite{k4}. Given the progress of AI-assisted programming, there's a significant opportunity to further apply these innovations to software testing and validation. By addressing the dual problems of increasing the bug detection rate and decreasing the bug rate in code generation, AI can play a pivotal role in transforming traditional testing paradigms.

\begin{figure*}
\centerline{\includegraphics[width=30pc]{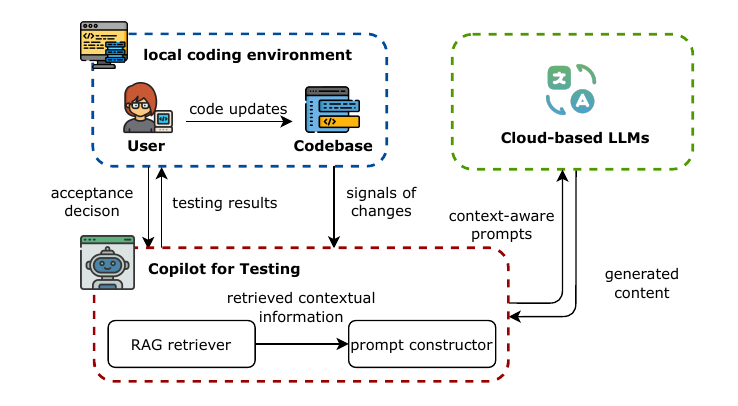}}
\caption{Architecture of \textit{Copilot for Testing}: It proactively observes the local coding environment, retrieves code context, and generates context-aware prompts to interact with cloud-based LLMs and provide bug fix suggestions synchronized with codebase updates, aiming at more efficient and effective software testing with higher accuracy and coverage.}\vspace*{-5pt}
\end{figure*}

We have previously launched and open-sourced \textit{Copilot for Xcode},\footnote{\tt \url{https://github.com/intitni/CopilotForXcode}} an AI-assisted programming tool for code auto-completion and Q\&A, which was later re-licensed and assimilated into GitHub.\footnote{\tt \url{https://github.blog/changelog/2024-10-29-github-copilot-code-completion-in-xcode-is-now-available-in-public-preview/}} Our initial system demonstrated the effectiveness of using contextual information to enhance the capabilities of LLMs in real-time code generation and problem-solving, which inspired us to extend the solution to software testing by transforming the challenge of generating code with fewer bugs into one of detecting and fixing bugs in the codebase.

This paper introduces {\it Copilot for Testing}, an AI-assisted testing system that provides bug detection, fix suggestions, and automated test case generation directly within the development environment, in sync with codebase updates. Our proposed testing methodology integrates a context-based RAG mechanism that interacts dynamically with local coding environments and retrieves contextual information as an enhancement to the LLMs prompts. This interaction not only allows the system to adapt and refine testing strategies in real-time, responding to ongoing changes within the codebase, but also improves the performance of automated testing in efficiency, accuracy, and coverage. The main contributions of this work include:
\begin{itemize}
    \item We propose a software testing methodology powered by a context-based RAG module, achieving a 31.2\% improvement in bug detection accuracy and a 12.6\% increase in critical test coverage compared to the baseline.  
    \item We develop \textit{Copilot for Testing}, an automated software testing system that seamlessly integrates with the development environment to deliver bug detection, fix suggestions, and test case generation in sync with codebase updates, achieving a 10.5\% incremental acceptance rate for code suggestions in our user studies.
\end{itemize}

\section{RELATED WORK}
\subsection{AI-Assisted Programming and Testing with Large Language Models}
The integration of AI, particularly through Large Language Models (LLMs), has revolutionized both programming and testing by automating and enhancing various tasks\cite{k5}. AI-assisted programming tools like \textit{GitHub Copilot} have demonstrated the potentials of LLMs in code completion and bug detection by understanding and generating code\cite{k6, k7, k8}. These dual problems—writing correct code and verifying code correctness—although distinct, share a common goal of reducing bugs and enhancing code quality. Our work extends this paradigm by offering a unified solution to both.

\subsection{Automated Software Testing}
Automated software testing plays a crucial role in modern software development practices. Recent advancements include test case generation with machine learning techniques which reduces manual design efforts\cite{k9}, automated test execution which can be integrated with continuous integration pipelines, and automated test results analysis which enhances decision-making processes regarding software quality\cite{k10}. Despite these advancements, challenges such as flaky tests and maintenance issues persist, highlighting the need for more robust testing methods and smarter maintenance strategies. Common metrics include accuracy, efficiency, and coverage\cite{k1}, which also guide our evaluation strategies.

\begin{figure*}
\centerline{\includegraphics[width=36pc]{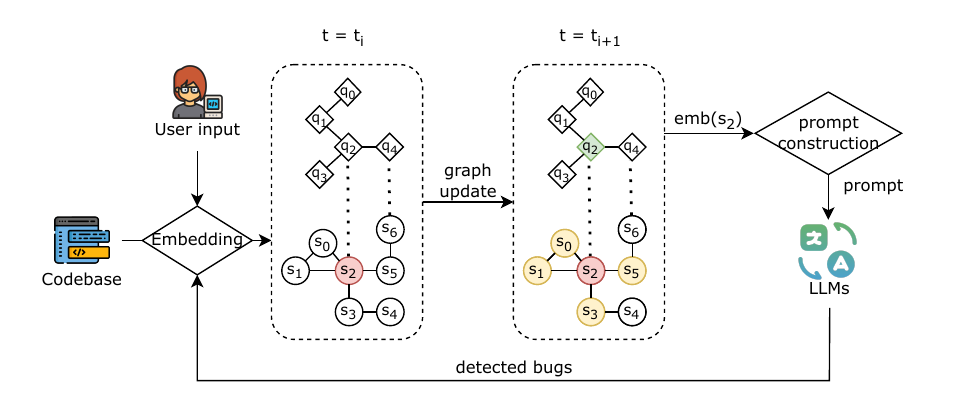}}
\caption{Overview of the context-based RAG module. The codebase is modeled as a graph, with individual nodes of code context embeddings. These embeddings are dynamically updated based on code changes initiated by the user and the outputs from LLMs. The toy example shows that when a change is made to node $s_2$, its embedding is updated, followed by updates to its neighboring nodes. The updated embeddings, which carry contextual information, are then utilized in prompt construction for the LLMs.}\vspace*{-5pt}
\end{figure*}

\subsection{Search-Based Software Engineering}
Search-Based Software Engineering (SBSE) utilizes optimization algorithms to solve software engineering problems, where solutions are systematically searched based on a defined fitness function and problem representation\cite{k11}. It has been successfully applied to various software engineering tasks including test case generation, software refactoring, and requirement engineering. SBSE's flexibility in defining fitness functions and representations makes it ideal for tailoring solutions to specific software engineering challenges, especially those involving large complex problem spaces. In our research, we model the problem within the SBSE framework and search for the optimal contextual information for the prompt construction.

\subsection{Retrieval Augmented Generation}
Retrieval Augmented Generation (RAG) combines retrieval mechanisms with generative models to enhance output accuracy and contextual relevance\cite{k12}. Originally developed for natural language processing, RAG has shown promise in software engineering for tasks like bug detection\cite{k13}. By integrating relevant code snippets or bug reports, RAG provides essential context to Large Language Models, improving the precision of generated solutions. Our approach leverages RAG to refine both programming and testing processes, ensuring AI-generated outputs are highly informed and effective.

\section{PROBLEM STATEMENT}
We position the research problem within the context of SBSE, where the goal is to find a solution that maximizes the fitness function based on the problem’s representation. In our work, we aim to enhance LLM-driven code generation by optimizing the contextual information provided. Here, the problem’s representation is defined as code context embeddings, while the fitness function reflects key metrics from automated testing, such as efficiency, accuracy, and coverage. This research is thus guided by the following key questions:

\textbf{RQ1}: How can we improve the performance of AI-assisted software testing by utilizing context-based RAG to enhance bug detection and code generation in real-time?

\textbf{RQ2}: How can we scale AI-assisted software testing to handle larger software projects while maintaining testing efficiency, accuracy, and coverage?

\section{METHODOLOGY}
We propose a code context-based AI-assisted testing system that synchronizes bug detection, fix suggestion, and test case generation with codebase updates, alongside the software engineer’s coding activities. The subsequent subsections detail the architecture and system flow.

\subsection{Architecture}
Figure 1 illustrates the architecture of the proposed system, which includes three components: the local coding environment that involves the user and the codebase, cloud-based LLMs, and \textit{Copilot for Testing} which includes the proposed RAG retriever and LLM prompt constructor. The local coding environment involves the user and the codebase which updates as the user edits code or accepts code suggestions. The external LLM API receives specific prompts and generates content accordingly. \textit{Copilot for Testing} proactively monitors changes in the codebase, retrieves contextual information with the context-based RAG retriever, provides context-aware prompts with the prompt constructor, and provides fix suggestions for detected bugs from LLMs' outputs.

Implemented as a plug-in for the coding IDE, \textit{Copilot for Testing} typically requires access permissions to the local codebase and user interactions. We demonstrate the implementation details of its application within Xcode in the later section.

\subsection{Context-Based RAG Retriever}
RAG techniques generally enhance content generation by providing an additional knowledge base. The objective of the proposed context-based RAG retriever is to deliver highly relevant content from the codebase to optimize code generation. It does so by retrieving contextual information that accurately reflects the current state of the local coding environment, thereby providing insightful context for the generator. In our approach, the retriever learns code context embeddings that capture multiple factors and dynamically adapts them based on real-time code changes and error patterns. These embeddings serve as structured representations of the evolving code context, helping to refine prompt construction for LLM-based testing and generation.

To model the codebase, we use graph-based representations, as shown in Figure 2. In this model, nodes represent code context embeddings, and edges denote connections. These embeddings incorporate several dynamically updated factors, including:

\begin{itemize}
    \item \textbf{File Path}---The relative and absolute paths in the filesystem, which provide spatial and organizational context.
    \item \textbf{Cursor Position}---The user's current focus within a file, indicating the most relevant areas for immediate context.
    \item \textbf{File Content}---Semantic information derived from the code within each file.
    \item \textbf{Bug Logs}---Historical bug reports and test outcomes, which indicate areas with higher defect frequency or complexity.
    \item \textbf{Graph Connectivity}---Metrics such as the number of neighbors and centrality, which reflect the complexity and interdependence of the code.
\end{itemize}

The weights of these factors are assigned based on empirical evaluation where different combinations of factors were tested for their impact on testing performance. For instance, we found that bug logs and cursor position provided strong indicators of relevance for immediate context, while graph connectivity was particularly useful for identifying critical file paths. 

The embeddings evolve continuously through information propagation within the graph, triggered by code edits, detected bugs, and test outcomes. Typically, this information flows from modified nodes to neighboring nodes along the edges, with diminishing weight to focus updates on a localized neighborhood, to keep the computational overhead manageable and the context learning accurate. Meanwhile, insights from LLM-generated test results and bug detections further refine the embeddings over time, creating a self-improving feedback loop that enhances testing precision.

Consequently, we frame the AI-assisted testing problem as optimizing the contextual information included in the prompts for LLMs to enhance testing effectiveness.

\subsection{Context-Aware Prompt Constructor} 
Following the RAG retriever, another key component of \textit{Copilot for Testing} is the LLM prompt constructor, which integrates retrieved content, user interaction history, and contextual information to generate effective prompts.

Typically, a prompt is structured in JSON format and comprises four main components:
\begin{enumerate}
\item \textbf{Context System Prompt} --- retrieved from the local code context
\item \textbf{Message History} --- an aggregated collection of past interactions between the user and the copilot
\item \textbf{Current Question} --- the current task to be executed, such as ``generate tests," ``execute tests," and ``analyze test results"
\item \textbf{Config System Prompt} --- overarching instructions, including model parameters, temperature, and mode settings
\end{enumerate}

The primary goal of the prompt constructor is to determine the optimal combination and ranking of these components to maximize prompt effectiveness.

\begin{figure*}
\centerline{\includegraphics[width=36pc]{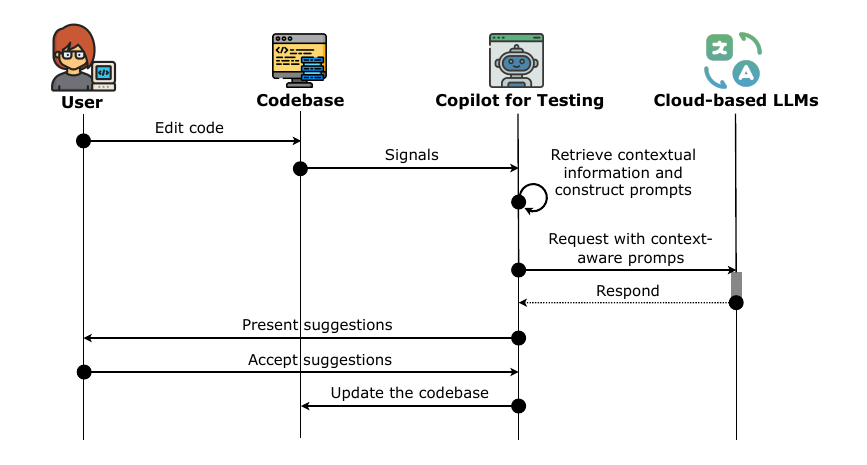}}
\caption{Sequence Diagram of {\it Copilot for Testing}. The sequence diagram illustrates the functionality of {\it Copilot for Testing} which enables real-time auto-synchronized testing through context-based RAG that leverages LLMs. {\it Copilot for Testing} receives notifications upon code updates, retrieves contextual information, and subsequently constructs prompts with the proposed RAG module. Upon receiving the suggestions, the user has the option to adopt the recommendations and directly apply the changes to the code base.}\vspace*{-5pt}
\end{figure*}

\subsection{System Flow}
Figure 3 illustrates the system flow of \textit{Copilot for Testing} in response to code changes made by the user. Upon detecting changes, the context-based RAG module updates the codebase graph and retrieves contextually relevant information for prompt construction. Based on the retrieved context, it presents detected bugs or suggested fixes to the user. When inconsistencies, errors, or anomalous patterns are identified, it prioritizes bug detection, providing targeted fixes along with contextual explanations. Otherwise, it emphasizes code completion and enhancement. The generated content is also utilized to update the graph node embeddings. The sensitivity of the code change listeners can be configured to prevent overly frequent updates, thereby conserving computational resources.

\textit{Copilot for Testing} is designed to automatically synchronize with local codebase changes, enabling all steps in the workflow to be executed in parallel as asynchronous jobs. This approach ensures scalable and efficient project-wide test coverage while maintaining a self-healing testing environment. The dynamic interaction within the system not only streamlines the testing process but also reduces the manual effort required from developers.

\subsection{Implementation Details}
We previously implemented \textit{Copilot for Xcode}, an AI-assisted programming tool for Xcode\cite{k14}. The project was unprecedented and challenging due to the restrictions in Apple's development environment, which generally blocks plugins. We overcame these limitations by leveraging the Accessibility API and network services to accurately track user focus points and detect real-time code edits, enabling a seamless user experience and contextually intelligent code generation. Building on this foundation, we extended the tool to implement \textit{Copilot for Testing} for automated testing, utilizing the contextual RAG-powered LLM capabilities for software testing. These extensions include modeling and maintaining embedded graphs for code repositories and customizing the context-aware prompts for bug detection, fix suggestions, and test case generation. 

We selected Xcode as our initial platform due to its highly restrictive sandbox environment, which presents significant challenges for retrieving and learning local code context. By successfully overcoming these challenges in Xcode, we demonstrate the robustness and adaptability of our approach. The framework’s design is inherently \textbf{platform-agnostic}, relying on modular components such as the RAG module and graph-based embeddings, which can be easily adapted to other IDEs with fewer restrictions. This ensures that expanding to other platforms, such as Visual Studio, IntelliJ, or Eclipse, will require minimal adjustments, primarily involving the integration of platform-specific APIs for context retrieval.

\section{Evaluation}
We evaluate the performance of \textit{Copilot for Testing} via both objective and subjective experiments, in comparison with the baseline model which does not leverage the context-based RAG module. Experiment results have proved the validity and usability of \textit{Copilot for Testing}, demonstrating its outstanding performance compared to the baseline. As aligned with common topline metrics in test automation$^{15}$, we assess the effectiveness of \textit{Copilot for Testing} from the following aspects.
\begin{itemize}
    \item {\it Accuracy}—Evaluated by the bug detection rate, emphasizing the method's ability to identify complex bugs involving multi-file interactions, thus demonstrating its precision in detecting subtle issues.
    \item {\it Efficiency}---Evaluated by the acceptance rate of the suggested fix for detected bugs or generated test cases since accepted generations directly imply saved engineering efforts. It is also understood from the subjective feedback from participants in the user study regarding their experiences in real projects.
    \item {\it Coverage}—Evaluated by the proportion of the codebase covered by automatically generated tests. We prioritize test coverage on critical code paths to optimize efficiency, rather than maximizing overall coverage, which could lead to unnecessary tests and increased costs, in order to focus on essential areas to the application's functionality
\end{itemize}

We utilized a database of known issues and test cases derived from open-source Swift projects and adapted C++ projects from the Software-artifact Infrastructure Repository.\footnote{\url{https://sir.csc.ncsu.edu/}} This provided a diverse set of challenges representative of real-world software development. Our evaluation included both objective experiments, to quantitatively measure the accuracy of the testing method, and subjective experiments, where software engineers evaluated the efficiency in practical scenarios.

\subsection{Objective Evaluation}
We conducted objective evaluations using a curated database of Swift and C++ projects from SIR, each containing known bugs (artificial faults called ``mutants") of varying complexity, to assess the \textbf{accuracy}, \textbf{overall coverage}, and \textbf{critical coverage} of the proposed testing methodology in comparison to the baseline model.

Following the standardized workflow of experimenting with SIR programs,\footnote{\url{https://github.com/jwlin/SIR-note/blob/master/workflow.md}} we executed the subject programs with their test cases and mutants, collecting fault-revealing and coverage data for tracking analysis. Accuracy was measured as the percentage of successfully detected bugs, while coverage was assessed as the percentage of code paths covered by generated tests. Additionally, we introduced ``critical coverage", which prioritizes high-impact code areas most relevant to system functionality. This ensures an efficiency-driven balance between exhaustive test coverage and practical maintainability.

\begin{table*}[t]
\centering
\caption{Evaluation Results: Proposed Model vs. Baseline Model}
\begin{tabular}{lccc}
\hline
\textbf{Metric}                     & \textbf{Proposed Model} & \textbf{Baseline Model} & \textbf{Change from Baseline} \\ \hline
Bug Detection Accuracy        & 85.3\%                       & 54.1\%  & +31.2\%                    \\ 
Overall Test Coverage          & 68.7\%                      & 70.0\%       & -1.3\%             \\ 
Critical Coverage              & 83.6\%                        & 71.0\%      & +12.6\%              \\ 
Cross-File Bug Detection       & 81.2\%                        & 49.0\%  & +32.2\%                  \\ 
Execution Time Per Bug    & 0.42 seconds                        & 0.68 seconds   & -                  \\ 
Suggestion Acceptance Rate    & 31.9\%                         & 21.4\%  & +10.5\%                  \\ \hline
\end{tabular}
\label{tab:objective_eval}
\end{table*}

The results are summarized in Table \ref{tab:objective_eval}. Our evaluation revealed that the proposed testing methodology achieved a \textbf{31.2\%} higher bug detection rate compared to the baseline model, excelling at identifying intricate bugs that involve cross-file dependencies with a \textbf{32.2\%} higher detection rate. While the overall test coverage was slightly lower (\textbf{-1.3\%}) than the baseline model, this reduction is a strategic trade-off, as the focus shifted to enhancing ``critical coverage'' which increased by \textbf{12.6\%}. To compute the ``critical coverage'' rate, we leveraged graph node embeddings which encompass factors like bug logs, change logs, and complexity of dependencies to represent the importance level of each code path and then measured the test coverage among the most important code paths. 

The contrasting results between overall and critical test coverage highlight a strategic tradeoff: by prioritizing critical code paths, the proposed system achieves higher efficiency and effectiveness in detecting complex, high-impact bugs, even with a marginal reduction in overall coverage. This tradeoff is justified by the significant improvements in bug detection rate and execution efficiency, as the overall execution time per detected bug was reduced from 0.68 seconds to 0.42 seconds. These results demonstrate that the slight reduction in overall coverage is outweighed by the substantial gains in critical coverage and system efficiency, making the approach highly valuable for developers who prioritize identifying critical bugs over exhaustive but less impactful testing.

The advanced performance of the proposed model in accuracy can be attributed to its dynamic adaptation to code changes and the deep contextual insights it provides. Compared to the baseline model, the proposed system’s use of contextual information enables a more nuanced understanding of the code structure, making it more adept at capturing subtle and complex bugs. The usage of graph-based embeddings offers higher information density, reducing the risk of distracting the LLM with less relevant data. 

\subsection{Subjective Evaluation}
To evaluate the efficiency and usability of \textit{Copilot for Testing}, we conducted a user study with 12 iOS developers. The study aimed to assess key metrics such as the acceptance rate of suggestions and gather qualitative feedback on the tool's practical applicability. Participants were divided into two groups: the test group using the proposed module and the control group using the baseline version. They were asked to complete testing-related tasks, designed to reflect common testing workflows, including debugging, generating test cases, and verifying the functionality of algorithms, UI elements, and database operations. The acceptance and rejection of code suggestions from both models were logged, and participants were also asked to evaluate the ease of use, compatibility with existing development practices, and overall impact on their testing workflows.

The acceptance rate was computed as the ratio of accepted suggestions to total suggestions triggered. Results show a \textbf{10.5\%} higher acceptance rate with the proposed model compared to the baseline. Additionally, participants reported a significant reduction in manual testing efforts, highlighting the value of automated test case generation. They also appreciated the system's ability to adapt quickly to code changes and provide timely feedback.

On the other hand, participants noted a steep learning curve during the initial setup of the testing module. They also observed slower response times during bulk operations, such as mass file generation or removal, which we attributed to the graph structure update process. When asked about desired features, participants expressed a strong interest in automated test execution with integrated results for bug detection and fixes, as well as support for multiple programming languages and development environments. These insights offer valuable directions for further enhancements and refinement of the system.

\section{FUTURE RESEARCH DIRECTIONS}
As we continue to refine and expand the capabilities of the proposed automated testing module, our research and development efforts will focus on several key areas:
\begin{enumerate}
    \item \textbf{End-to-End Parameter Tuning:} In the current testing system, many parameters, such as the weights assigned to different sources of information for node embedding and the ranking of components for prompt construction, are configured based on optimal values determined through trial and error. In future studies, we aim to incorporate these parameter settings into an end-to-end dynamic training procedure, enabling the system to learn and optimize these values automatically.
    \item \textbf{User Experience Improvements:} We plan to enhance the user experience by implementing algorithmic optimizations to reduce loading and pending times, ensuring smoother performance when working with large codebases. For example, we will leverage caching mechanisms for executed test results to eliminate redundant work and improve efficiency. Additionally, to reduce the learning curve, we will introduce in-UI tips that provide step-by-step guidance, helping users navigate the system more effectively during setup and daily use.
    \item \textbf{Expansion to Broader Platforms:} We aim to extend the tool's compatibility to multiple development environments and programming languages beyond Xcode and Swift. The proposed RAG module and graph-based embeddings are language-agnostic and platform-independent, making them easily generalizable to other less restrictive platforms. For example, integrating the framework into another IDE would primarily involve adapting the context retrieval mechanism to leverage that platform's local API, while the core RAG and testing logic would remain unchanged. This expansion will enhance the proposed solution’s accessibility and appeal to a broader range of developers, thereby improving its usability and adoption.
\end{enumerate}

\section{CONCLUSION}
In this study, we proposed a software testing solution with context-based RAG to enable bug detection, fix suggestions, and test case generation synchronized with coding, as an effective extension from AI-assisted code generation to software testing. We validated its improved performance in accuracy, efficiency, and coverage over the baseline through both objective and subjective experiments. These results demonstrate the potential of leveraging code contextual information to enhance LLMs for code generation and project development, helping to keep pace with the increasing complexity and scale of modern software systems, while establishing a foundation for future innovations in AI-assisted development tools.

\section*{Acknowledgment}
This research was supported by the Singapore Ministry of Education Academic Research Fund under Grant RG91/22.

\bibliographystyle{unsrtnat}
\bibliography{references}  






\end{document}